\documentstyle[11pt,paspconf]{article}
\def\fileversion{v1.13}%
\def\filedate{6.5.93}%
\edef\epsfigRestoreAt{\catcode`@=\number\catcode`@\relax}%
\catcode`\@=11\relax
\immediate\write16{Document style option `epsfig', \fileversion\space
<\filedate> (edited by SPQR)}%
\newcount\EPS@Height
\newcount\EPS@Width
\newcount\EPS@xscale
\newcount\EPS@yscale
\def\psfigdriver#1{%
  \bgroup\edef\next{\def\noexpand\tempa{#1}}%
    \uppercase\expandafter{\next}%
    \def\LN{DVITOLN03}%
    \def\DVItoPS{DVITOPS}%
    \def\DVIPS{DVIPS}%
    \def\emTeX{EMTEX}%
    \def\OzTeX{OZTEX}%
    \def\Textures{TEXTURES}%
    \global\chardef\fig@driver=0
    \ifx\tempa\LN
        \global\chardef\fig@driver=0\fi
    \ifx\tempa\DVItoPS
        \global\chardef\fig@driver=1\fi
    \ifx\tempa\DVIPS
        \global\chardef\fig@driver=2\fi
    \ifx\tempa\emTeX
        \global\chardef\fig@driver=3\fi
    \ifx\tempa\OzTeX
        \global\chardef\fig@driver=4\fi
    \ifx\tempa\Textures
        \global\chardef\fig@driver=5\fi
  \egroup
\def\psfig@start{}%
\def\psfig@end{}%
\def\epsfig@gofer{}%
\ifcase\fig@driver
\typeout{WARNING! ****
 no specials for LN03 psfig}%
\or % case 1: dvitops
\def\psfig@start{}%
\def\psfig@end{\special{dvitops: import \@p@sfilefinal \space
\@p@swidth sp \space \@p@sheight sp \space fill}%
\if@clip \typeout{Clipping not supported}\fi
\if@angle \typeout{Rotating not supported}\fi
}%
\let\epsfig@gofer\psfig@end
\or %case2 dvips
\def\psfig@start{\special{ps::[begin]  \@p@swidth \space \@p@sheight \space%
        \@p@sbbllx \space \@p@sbblly \space%
        \@p@sbburx \space \@p@sbbury \space%
        startTexFig \space }%
        \if@angle
                \special {ps:: \@p@sangle \space rotate \space}
        \fi
        \if@clip
                \if@verbose
                        \typeout{(clipped to BB) }%
                \fi
                \special{ps:: doclip \space }%
        \fi
        \special{ps: plotfile \@p@sfilefinal \space }%
        \special{ps::[end] endTexFig \space }%
}%
\def\psfig@end{}%
\def\epsfig@gofer{\if@clip
                        \if@verbose
                           \typeout{(clipped to BB)}%
                        \fi
                        \epsfclipon
                  \fi
                  \epsfsetgraph{\@p@sfilefinal}%
}%
\or % case 3, emTeX
\typeout{WARNING. You must have a .bb info file with the Bounding Box
  of the pcx file}%
\def\psfig@start{}%
\def\psfig@end{\typeout{pcx import of \@p@sfilefinal}%
\if@clip \typeout{Clipping not supported}\fi
\if@angle \typeout{Rotating not supported}\fi
\raisebox{\@p@srheight true sp}{\special{em: graph \@p@sfilefinal}}}%
\def\epsfig@gofer{}%
\or % case 4, OzTeX
\def\psfig@start{}%
\def\psfig@end{%
\EPS@Width\@p@swidth
\EPS@Height\@p@sheight
\divide\EPS@Width by 65781  % convert sp to bp
\divide\EPS@Height by 65781
\special{epsf=\@p@sfilefinal
\space
width=\the\EPS@Width
\space
height=\the\EPS@Height
}%
\if@clip \typeout{Clipping not supported}\fi
\if@angle \typeout{Rotating not supported}\fi
}%
\let\epsfig@gofer\psfig@end
\or % case 5, Textures
\def\psfig@end{\if@clip
                        \if@verbose
                           \typeout{(clipped to BB)}%
                        \fi
                        \epsfclipon
                  \fi
\special{illustration \@p@sfilefinal\space scaled \the\EPS@xscale}%
}%
\def\psfig@start{}%
\let\epsfig\psfig
\else
\typeout{WARNING. *** unknown  driver - no psfig}%
\fi
}%
\newdimen\ps@dimcent
\ifx\undefined\fbox
\newdimen\fboxrule
\newdimen\fboxsep
\newdimen\ps@tempdima
\newbox\ps@tempboxa
\long\def\fbox#1{\leavevmode\setbox\ps@tempboxa\hbox{#1}\ps@tempdima\fboxrule
    \advance\ps@tempdima \fboxsep \advance\ps@tempdima \dp\ps@tempboxa
   \hbox{\lower \ps@tempdima\hbox
  {\vbox{\hrule height \fboxrule
          \hbox{\vrule width \fboxrule \hskip\fboxsep
          \vbox{\vskip\fboxsep \box\ps@tempboxa\vskip\fboxsep}\hskip
                 \fboxsep\vrule width \fboxrule}%
                 \hrule height \fboxrule}}}}%
\fi
\ifx\@ifundefined\undefined
\long\def\@ifundefined#1#2#3{\expandafter\ifx\csname
  #1\endcsname\relax#2\else#3\fi}%
\fi
\@ifundefined{typeout}%
{\gdef\typeout#1{\immediate\write\sixt@@n{#1}}}%
{\relax}%
\@ifundefined{epsfig}{}{\typeout{EPSFIG --- already loaded}\endinput}%
\@ifundefined{epsfbox}{\input epsf}{}%
\ifx\undefined\@latexerr
        \newlinechar`\^^J
        \def\@spaces{\space\space\space\space}%
        \def\@latexerr#1#2{%
        \edef\@tempc{#2}\expandafter\errhelp\expandafter{\@tempc}%
        \typeout{Error. \space see a manual for explanation.^^J
         \space\@spaces\@spaces\@spaces Type \space H <return> \space for
         immediate help.}\errmessage{#1}}%
\fi
\def\@whattodo{You tried to include a PostScript figure which
cannot be found^^JIf you press return to carry on anyway,^^J
The failed name will be printed in place of the figure.^^J
or type X to quit}%
\def\@whattodobb{You tried to include a PostScript figure which
has no^^Jbounding box, and you supplied none.^^J
If you press return to carry on anyway,^^J
The failed name will be printed in place of the figure.^^J
or type X to quit}%
\def\@nnil{\@nil}%
\def\@empty{}%
\def\@psdonoop#1\@@#2#3{}%
\def\@psdo#1:=#2\do#3{\edef\@psdotmp{#2}\ifx\@psdotmp\@empty \else
    \expandafter\@psdoloop#2,\@nil,\@nil\@@#1{#3}\fi}%
\def\@psdoloop#1,#2,#3\@@#4#5{\def#4{#1}\ifx #4\@nnil \else
       #5\def#4{#2}\ifx #4\@nnil \else#5\@ipsdoloop #3\@@#4{#5}\fi\fi}%
\def\@ipsdoloop#1,#2\@@#3#4{\def#3{#1}\ifx #3\@nnil
       \let\@nextwhile=\@psdonoop \else
      #4\relax\let\@nextwhile=\@ipsdoloop\fi\@nextwhile#2\@@#3{#4}}%
\def\@tpsdo#1:=#2\do#3{\xdef\@psdotmp{#2}\ifx\@psdotmp\@empty \else
    \@tpsdoloop#2\@nil\@nil\@@#1{#3}\fi}%
\def\@tpsdoloop#1#2\@@#3#4{\def#3{#1}\ifx #3\@nnil
       \let\@nextwhile=\@psdonoop \else
      #4\relax\let\@nextwhile=\@tpsdoloop\fi\@nextwhile#2\@@#3{#4}}%
\long\def\epsfaux#1#2:#3\\{\ifx#1\epsfpercent
   \def\testit{#2}\ifx\testit\epsfbblit
        \@atendfalse
        \epsf@atend #3 . \\%
        \if@atend
           \if@verbose
                \typeout{epsfig: found `(atend)'; continuing search}%
           \fi
        \else
                \epsfgrab #3 . . . \\%
                \epsffileokfalse\global\no@bbfalse
                \global\epsfbbfoundtrue
        \fi
   \fi\fi}%
\def\epsf@atendlit{(atend)}
\def\epsf@atend #1 #2 #3\\{%
   \def\epsf@tmp{#1}\ifx\epsf@tmp\empty
      \epsf@atend #2 #3 .\\\else
   \ifx\epsf@tmp\epsf@atendlit\@atendtrue\fi\fi}%

\chardef\trig@letter = 11
\chardef\other = 12

\newif\ifdebug %%% turn me on to see TeX hard at work ...
\newif\ifc@mpute %%% don't need to compute some values
\newif\if@atend
\c@mputetrue % but assume that we do

\let\then = \relax
\def\r@dian{pt }%
\let\r@dians = \r@dian
\let\dimensionless@nit = \r@dian
\let\dimensionless@nits = \dimensionless@nit
\def\internal@nit{sp }%
\let\internal@nits = \internal@nit
\newif\ifstillc@nverging
\def \Mess@ge #1{\ifdebug \then \message {#1} \fi}%

{ %%% Things that need abnormal catcodes %%%
        \catcode `\@ = \trig@letter
        \gdef \nodimen {\expandafter \n@dimen \the \dimen}%
        \gdef \term #1 #2 #3%
               {\edef \t@ {\the #1}%%% freeze parameter 1 (count, by value)
                \edef \t@@ {\expandafter \n@dimen \the #2\r@dian}%
                \t@rm {\t@} {\t@@} {#3}%
               }%
        \gdef \t@rm #1 #2 #3%
               {{%
                \count 0 = 0
                \dimen 0 = 1 \dimensionless@nit
                \dimen 2 = #2\relax
                \Mess@ge {Calculating term #1 of \nodimen 2}%
                \loop
                \ifnum  \count 0 < #1
                \then   \advance \count 0 by 1
                        \Mess@ge {Iteration \the \count 0 \space}%
                        \Multiply \dimen 0 by {\dimen 2}%
                        \Mess@ge {After multiplication, term = \nodimen 0}%
                        \Divide \dimen 0 by {\count 0}%
                        \Mess@ge {After division, term = \nodimen 0}%
                \repeat
                \Mess@ge {Final value for term #1 of
                                \nodimen 2 \space is \nodimen 0}%
                \xdef \Term {#3 = \nodimen 0 \r@dians}%
                \aftergroup \Term
               }}%
        \catcode `\p = \other
        \catcode `\t = \other
        \gdef \n@dimen #1pt{#1} %%% throw away the ``pt''
}%

\def \Divide #1by #2{\divide #1 by #2} %%% just a synonym

\def \Multiply #1by #2%%% allows division of a dimen by a dimen
       {{%%% should really freeze parameter 2 (dimen, passed by value)
        \count 0 = #1\relax
        \count 2 = #2\relax
        \count 4 = 65536
        \Mess@ge {Before scaling, count 0 = \the \count 0 \space and
                        count 2 = \the \count 2}%
        \ifnum  \count 0 > 32767 %%% do our best to avoid overflow
        \then   \divide \count 0 by 4
                \divide \count 4 by 4
        \else   \ifnum  \count 0 < -32767
                \then   \divide \count 0 by 4
                        \divide \count 4 by 4
                \else
                \fi
        \fi
        \ifnum  \count 2 > 32767 %%% while retaining reasonable accuracy
        \then   \divide \count 2 by 4
                \divide \count 4 by 4
        \else   \ifnum  \count 2 < -32767
                \then   \divide \count 2 by 4
                        \divide \count 4 by 4
                \else
                \fi
        \fi
        \multiply \count 0 by \count 2
        \divide \count 0 by \count 4
        \xdef \product {#1 = \the \count 0 \internal@nits}%
        \aftergroup \product
       }}%

\def\r@duce{\ifdim\dimen0 > 90\r@dian \then   % sin(x) = sin(180-x)
                \multiply\dimen0 by -1
                \advance\dimen0 by 180\r@dian
                \r@duce
            \else \ifdim\dimen0 < -90\r@dian \then  % sin(x) = sin(360+x)
                \advance\dimen0 by 360\r@dian
                \r@duce
                \fi
            \fi}%

\def\Sine#1%
       {{%
        \dimen 0 = #1 \r@dian
        \r@duce
        \ifdim\dimen0 = -90\r@dian \then
           \dimen4 = -1\r@dian
           \c@mputefalse
        \fi
        \ifdim\dimen0 = 90\r@dian \then
           \dimen4 = 1\r@dian
           \c@mputefalse
        \fi
        \ifdim\dimen0 = 0\r@dian \then
           \dimen4 = 0\r@dian
           \c@mputefalse
        \fi
        \ifc@mpute \then
                \divide\dimen0 by 180
                \dimen0=3.141592654\dimen0
                \dimen 2 = 3.1415926535897963\r@dian %%% a well-known constant
                \divide\dimen 2 by 2 %%% we only deal with -pi/2 : pi/2
                \Mess@ge {Sin: calculating Sin of \nodimen 0}%
                \count 0 = 1 %%% see power-series expansion for sine
                \dimen 2 = 1 \r@dian %%% ditto
                \dimen 4 = 0 \r@dian %%% ditto
                \loop
                        \ifnum  \dimen 2 = 0 %%% then we've done
                        \then   \stillc@nvergingfalse
                        \else   \stillc@nvergingtrue
                        \fi
                        \ifstillc@nverging %%% then calculate next term
                        \then   \term {\count 0} {\dimen 0} {\dimen 2}%
                                \advance \count 0 by 2
                                \count 2 = \count 0
                                \divide \count 2 by 2
                                \ifodd  \count 2 %%% signs alternate
                                \then   \advance \dimen 4 by \dimen 2
                                \else   \advance \dimen 4 by -\dimen 2
                                \fi
                \repeat
        \fi
                        \xdef \sine {\nodimen 4}%
       }}%

\def\Cosine#1{\ifx\sine\UnDefined\edef\Savesine{\relax}\else
                             \edef\Savesine{\sine}\fi
        {\dimen0=#1\r@dian\multiply\dimen0 by -1
         \advance\dimen0 by 90\r@dian
         \Sine{\nodimen 0}%
         \xdef\cosine{\sine}%
         \xdef\sine{\Savesine}}}
\def\psdraft{\def\@psdraft{0}}%
\def\psfull{\def\@psdraft{1}}%
\psfull
\newif\if@scalefirst
\def\psscalefirst{\@scalefirsttrue}%
\def\psrotatefirst{\@scalefirstfalse}%
\psrotatefirst
\newif\if@draftbox
\def\psnodraftbox{\@draftboxfalse}%
\@draftboxtrue
\newif\if@noisy
\@noisyfalse
\newif\ifno@bb
\newif\if@bbllx
\newif\if@bblly
\newif\if@bburx
\newif\if@bbury
\newif\if@height
\newif\if@width
\newif\if@rheight
\newif\if@rwidth
\newif\if@angle
\newif\if@clip
\newif\if@verbose
\newif\if@prologfile
\def\@p@@sprolog#1{\@prologfiletrue\def\@prologfileval{#1}}%
\def\@p@@sclip#1{\@cliptrue}%
\newif\ifepsfig@dos  % only single suffix possible
\def\epsfigdos{\epsfig@dostrue}%
\epsfig@dosfalse
\newif\ifuse@psfig
\def\ParseName#1{\expandafter\@Parse#1}%
\def\@Parse#1.#2:{\gdef\BaseName{#1}\gdef\FileType{#2}}%

\def\@p@@sfile#1{%
\ifepsfig@dos
   \ParseName{#1:}%
\else
   \gdef\BaseName{#1}\gdef\FileType{}%
\fi
\def\@p@sfile{NO FILE: #1}%
\def\@p@sfilefinal{NO FILE: #1}%
        \openin1=#1
        \ifeof1\closein1
                \openin1=\BaseName.bb
                        \ifeof1\closein1
                                \if@bbllx\if@bblly\if@bburx\if@bbury
                                        \def\@p@sfile{#1}%
                                        \def\@p@sfilefinal{#1}%
                                        \fi\fi\fi
                                \else
                                        \@latexerr{ERROR.
PostScript file #1 not found}\@whattodo
                                        \@p@@sbbllx{100bp}%
                                        \@p@@sbblly{100bp}%
                                        \@p@@sbburx{200bp}%
                                        \@p@@sbbury{200bp}%
                                        \psdraft
                                \fi
                        \else
                                \closein1%
                                \edef\@p@sfile{\BaseName.bb}%
                                \typeout{using BB from \@p@sfile}%
                                \ifnum\fig@driver=3
                                  \edef\@p@sfilefinal{\BaseName.pcx}%
                                \else
                                 \ifepsfig@dos
                                 \edef\@p@sfilefinal{"`uncompress
                                   < \BaseName.Z"}%
                                \else
                                \edef\@p@sfilefinal{"`zcat `texfind
                                  #1.Z`"}%
                                \fi
                                \fi
                        \fi
        \else\closein1
                    \edef\@p@sfile{#1}%
                    \edef\@p@sfilefinal{#1}%
        \fi%
}%
\let\@p@@sfigure\@p@@sfile
\def\@p@@sbbllx#1{%
				            \@bbllxtrue
                \ps@dimcent=#1
                \edef\@p@sbbllx{\number\ps@dimcent}%
                \divide\ps@dimcent by65536
                \global\edef\epsfllx{\number\ps@dimcent}%
}%
\def\@p@@sbblly#1{%
                \@bbllytrue
                \ps@dimcent=#1
                \edef\@p@sbblly{\number\ps@dimcent}%
                \divide\ps@dimcent by65536
                \global\edef\epsflly{\number\ps@dimcent}%
}%
\def\@p@@sbburx#1{%
                \@bburxtrue
                \ps@dimcent=#1
                \edef\@p@sbburx{\number\ps@dimcent}%
                \divide\ps@dimcent by65536
                \global\edef\epsfurx{\number\ps@dimcent}%
}%
\def\@p@@sbbury#1{%
                \@bburytrue
                \ps@dimcent=#1
                \edef\@p@sbbury{\number\ps@dimcent}%
                \divide\ps@dimcent by65536
                \global\edef\epsfury{\number\ps@dimcent}%
}%
\def\@p@@sheight#1{%
                \@heighttrue
                \global\epsfysize=#1
                \ps@dimcent=#1
                \edef\@p@sheight{\number\ps@dimcent}%
}%
\def\@p@@swidth#1{%
                \@widthtrue
                \global\epsfxsize=#1
                \ps@dimcent=#1
                \edef\@p@swidth{\number\ps@dimcent}%
}%
\def\@p@@srheight#1{%
                \@rheighttrue\use@psfigtrue
                \ps@dimcent=#1
                \edef\@p@srheight{\number\ps@dimcent}%
}%
\def\@p@@srwidth#1{%
                \@rwidthtrue\use@psfigtrue
                \ps@dimcent=#1
                \edef\@p@srwidth{\number\ps@dimcent}%
}%
\def\@p@@sangle#1{%
                \use@psfigtrue
                \@angletrue
                \edef\@p@sangle{#1}%
}%
\def\@p@@ssilent#1{%
                \@verbosefalse
}%
\def\@p@@snoisy#1{%
                \@verbosetrue
}%
\def\@cs@name#1{\csname #1\endcsname}%
\def\@setparms#1=#2,{\@cs@name{@p@@s#1}{#2}}%
\def\ps@init@parms{%
                \@bbllxfalse \@bbllyfalse
                \@bburxfalse \@bburyfalse
                \@heightfalse \@widthfalse
                \@rheightfalse \@rwidthfalse
                \def\@p@sbbllx{}\def\@p@sbblly{}%
                \def\@p@sbburx{}\def\@p@sbbury{}%
                \def\@p@sheight{}\def\@p@swidth{}%
                \def\@p@srheight{}\def\@p@srwidth{}%
                \def\@p@sangle{0}%
                \def\@p@sfile{}%
                \use@psfigfalse
                \@prologfilefalse
                \def\@sc{}%
                \if@noisy
                        \@verbosetrue
                \else
                        \@verbosefalse
                \fi
                \@clipfalse
}%
\def\parse@ps@parms#1{%
                \@psdo\@psfiga:=#1\do
                   {\expandafter\@setparms\@psfiga,}%
\if@prologfile
\fi
}%
\def\bb@missing{%
        \if@verbose
            \typeout{psfig: searching \@p@sfile \space  for bounding box}%
        \fi
        \epsfgetbb{\@p@sfile}%
        \ifepsfbbfound
            \ps@dimcent=\epsfllx bp\edef\@p@sbbllx{\number\ps@dimcent}%
            \ps@dimcent=\epsflly bp\edef\@p@sbblly{\number\ps@dimcent}%
            \ps@dimcent=\epsfurx bp\edef\@p@sbburx{\number\ps@dimcent}%
            \ps@dimcent=\epsfury bp\edef\@p@sbbury{\number\ps@dimcent}%
        \else
            \epsfbbfoundfalse
        \fi
}
\newdimen\p@intvaluex
\newdimen\p@intvaluey
\def\rotate@#1#2{{\dimen0=#1 sp\dimen1=#2 sp
                  \global\p@intvaluex=\cosine\dimen0
                  \dimen3=\sine\dimen1
                  \global\advance\p@intvaluex by -\dimen3
                  \global\p@intvaluey=\sine\dimen0
                  \dimen3=\cosine\dimen1
                  \global\advance\p@intvaluey by \dimen3
                  }}%
\def\compute@bb{%
                \epsfbbfoundfalse
                \if@bbllx\epsfbbfoundtrue\fi
                \if@bblly\epsfbbfoundtrue\fi
                \if@bburx\epsfbbfoundtrue\fi
                \if@bbury\epsfbbfoundtrue\fi
                \ifepsfbbfound\else\bb@missing\fi
                \ifepsfbbfound\else
                \@latexerr{ERROR. cannot locate BoundingBox}\@whattodobb
                        \@p@@sbbllx{100bp}%
                        \@p@@sbblly{100bp}%
                        \@p@@sbburx{200bp}%
                        \@p@@sbbury{200bp}%
                        \no@bbtrue
                        \psdraft
                \fi
                \count203=\@p@sbburx
                \count204=\@p@sbbury
                \advance\count203 by -\@p@sbbllx
                \advance\count204 by -\@p@sbblly
                \edef\ps@bbw{\number\count203}%
                \edef\ps@bbh{\number\count204}%
                 \edef\@bbw{\number\count203}%
                \edef\@bbh{\number\count204}%
               \if@angle
                        \Sine{\@p@sangle}\Cosine{\@p@sangle}%

{\ps@dimcent=\maxdimen\xdef\r@p@sbbllx{\number\ps@dimcent}%

\xdef\r@p@sbblly{\number\ps@dimcent}%

\xdef\r@p@sbburx{-\number\ps@dimcent}%

\xdef\r@p@sbbury{-\number\ps@dimcent}}%
                        \def\minmaxtest{%
                           \ifnum\number\p@intvaluex<\r@p@sbbllx
                              \xdef\r@p@sbbllx{\number\p@intvaluex}\fi
                           \ifnum\number\p@intvaluex>\r@p@sbburx
                              \xdef\r@p@sbburx{\number\p@intvaluex}\fi
                           \ifnum\number\p@intvaluey<\r@p@sbblly
                              \xdef\r@p@sbblly{\number\p@intvaluey}\fi
                           \ifnum\number\p@intvaluey>\r@p@sbbury
                              \xdef\r@p@sbbury{\number\p@intvaluey}\fi
                           }%
                        \rotate@{\@p@sbbllx}{\@p@sbblly}%
                        \minmaxtest
                        \rotate@{\@p@sbbllx}{\@p@sbbury}%
                        \minmaxtest
                        \rotate@{\@p@sbburx}{\@p@sbblly}%
                        \minmaxtest
                        \rotate@{\@p@sbburx}{\@p@sbbury}%
                        \minmaxtest

\edef\@p@sbbllx{\r@p@sbbllx}\edef\@p@sbblly{\r@p@sbblly}%

\edef\@p@sbburx{\r@p@sbburx}\edef\@p@sbbury{\r@p@sbbury}%
                \fi
                \count203=\@p@sbburx
                \count204=\@p@sbbury
                \advance\count203 by -\@p@sbbllx
                \advance\count204 by -\@p@sbblly
                \edef\@bbw{\number\count203}%
                \edef\@bbh{\number\count204}%
}%
\def\in@hundreds#1#2#3{\count240=#2 \count241=#3
                     \count100=\count240        % 100 is first digit #2/#3
                     \divide\count100 by \count241
                     \count101=\count100
                     \multiply\count101 by \count241
                     \advance\count240 by -\count101
                     \multiply\count240 by 10
                     \count101=\count240        %101 is second digit of #2/#3
                     \divide\count101 by \count241
                     \count102=\count101
                     \multiply\count102 by \count241
                     \advance\count240 by -\count102
                     \multiply\count240 by 10
                     \count102=\count240        % 102 is the third digit
                     \divide\count102 by \count241
                     \count200=#1\count205=0
                     \count201=\count200
                        \multiply\count201 by \count100
                        \advance\count205 by \count201
                     \count201=\count200
                        \divide\count201 by 10
                        \multiply\count201 by \count101
                        \advance\count205 by \count201
                     \count201=\count200
                        \divide\count201 by 100
                        \multiply\count201 by \count102
                        \advance\count205 by \count201
                     \edef\@result{\number\count205}%
}%
\def\compute@wfromh{%
                \in@hundreds{\@p@sheight}{\@bbw}{\@bbh}%
                \edef\@p@swidth{\@result}%
}%
\def\compute@hfromw{%
                \in@hundreds{\@p@swidth}{\@bbh}{\@bbw}%
                \edef\@p@sheight{\@result}%
}%
\def\compute@handw{%
                \if@height
                        \if@width
                        \else
                                \compute@wfromh
                        \fi
                \else
                        \if@width
                                \compute@hfromw
                        \else
                                \edef\@p@sheight{\@bbh}%
                                \edef\@p@swidth{\@bbw}%
                        \fi
                \fi
}%
\def\compute@resv{%
                \if@rheight \else \edef\@p@srheight{\@p@sheight} \fi
                \if@rwidth \else \edef\@p@srwidth{\@p@swidth} \fi
}%
\def\compute@sizes{%
        \if@scalefirst\if@angle
        \if@width
           \in@hundreds{\@p@swidth}{\@bbw}{\ps@bbw}%
           \edef\@p@swidth{\@result}%
        \fi
        \if@height
           \in@hundreds{\@p@sheight}{\@bbh}{\ps@bbh}%
           \edef\@p@sheight{\@result}%
        \fi
        \fi\fi
        \compute@handw
        \compute@resv
					           \EPS@Width=\@bbw
																\divide\EPS@Width by 1000
   												 \EPS@xscale=\@p@swidth \divide \EPS@xscale by \EPS@Width
					           \EPS@Height=\@bbh
																\divide\EPS@Height by 1000
   												 \EPS@yscale=\@p@sheight \divide \EPS@yscale by\EPS@Height
  \ifnum\EPS@xscale>\EPS@yscale\EPS@xscale=\EPS@yscale\fi
}
\def\psfig{\begingroup\@minisanitize\@@@psfig}
\def\epsfig{\begingroup\@minisanitize\@@@epsfig}

\def\@minisanitize{\@makeother\_\@makeother\:\@makeother\.\@makeother\$}

\def\@@@psfig#1{\vbox {%
        \ps@init@parms
        \parse@ps@parms{#1}%
        \ifnum\@psdraft=1
                \typeout{[\@p@sfilefinal]}%
                \if@verbose
                        \typeout{epsfig: using PSFIG macros}%
                \fi
                \psfig@method
        \else
                \epsfig@draft
        \fi
}
\endgroup
}%

\def\@@@epsfig#1{\vbox {%
        \ps@init@parms
        \parse@ps@parms{#1}%
        \ifnum\@psdraft=1
          \if@angle\use@psfigtrue\fi
          {\ifnum\fig@driver=1\global\use@psfigtrue\fi}%
          {\ifnum\fig@driver=3\global\use@psfigtrue\fi}%
          {\ifnum\fig@driver=4\global\use@psfigtrue\fi}%
          {\ifnum\fig@driver=5\global\use@psfigtrue\fi}%
                \ifuse@psfig
                        \if@verbose
                                \typeout{epsfig: using PSFIG macros}%
                        \fi
                        \psfig@method
                \else
                        \if@verbose
                                \typeout{epsfig: using EPSF macros}%
                        \fi
                        \epsf@method
                \fi
        \else
                \epsfig@draft
        \fi
}
\endgroup
}%

\def\epsf@method{%
        \epsfbbfoundfalse
        \if@bbllx\epsfbbfoundtrue\fi
        \if@bblly\epsfbbfoundtrue\fi
        \if@bburx\epsfbbfoundtrue\fi
        \if@bbury\epsfbbfoundtrue\fi
        \ifepsfbbfound\else\epsfgetbb{\@p@sfile}\fi
        \ifepsfbbfound
           \typeout{<\@p@sfilefinal>}%
           \epsfig@gofer
        \else
          \@latexerr{ERROR - Cannot locate BoundingBox}\@whattodobb
          \@p@@sbbllx{100bp}%
          \@p@@sbblly{100bp}%
          \@p@@sbburx{200bp}%
          \@p@@sbbury{200bp}%
                \count203=\@p@sbburx
                \count204=\@p@sbbury
                \advance\count203 by -\@p@sbbllx
                \advance\count204 by -\@p@sbblly
                \edef\@bbw{\number\count203}%
                \edef\@bbh{\number\count204}%
          \compute@sizes
          \epsfig@@draft
       \fi
}%
\def\psfig@method{%
        \compute@bb
        \ifepsfbbfound
          \compute@sizes
          \psfig@start
          \vbox to \@p@srheight true sp{\hbox to \@p@srwidth true
            sp{\hss}\vss\psfig@end}%
        \else
           \epsfig@draft
        \fi
}%
\def\epsfig@draft{\compute@bb\compute@sizes\epsfig@@draft}%
\def\epsfig@@draft{%
\typeout{<(draft only) \@p@sfilefinal>}%
\if@draftbox
        \hbox{\fbox{\vbox to \@p@srheight true sp{%
        \vss\hbox to \@p@srwidth true sp{ \hss
           {\tt\@p@sfilefinal}
                          \hss }\vss
        }}}%
\else
        \vbox to \@p@srheight true sp{%
        \vss\hbox to \@p@srwidth true sp{\hss}\vss}%
\fi
}%
\psfigdriver{dvips}%
\epsfigRestoreAt

\begin{document}
\title{Cepheids, supernovae and the value of $H_0$}
\author{M.A. Hendry}
\affil{Astronomy Centre, University of Sussex, Falmer, BN1 9QH, UK}
\author{S.M. Kanbur}
\affil{Dept of Physics and Astronomy, University of Glasgow, 
Glasgow G12 8QQ, UK}

\begin{abstract}
The calibration of the Type Ia supernova distances using HST observations
of Cepheid variables is discussed. A new maximum likelihood method of
calibration is applied to derive the PL relation for a composite sample
of Cepheids in the LMC and in the SNIa host galaxies NGC5253 and IC4182. 
Our results show that the calibration of the Cepheid
PL relation is robust both to sampling error and to luminosity and period
selection effects. Hence, the outstanding uncertainty in deriving estimates of
$H_0$ from SNIa remains the dispersion of the SNIa luminosity function, and
not unresolved systematic errors in its Cepheid calibration.
\end{abstract}


\section{Introduction}

Type Ia supernovae (henceforth SNIa) have long been regarded as useful
cosmological distance indicators because they are observable to large
velocity distances and their luminosity at
maximum light displays a small intrinsic dispersion. In e.g. Sandage \&
Tammann (1993) the Hubble diagram of 34 SNIa in or beyond the Virgo cluster
was found to have an observed V band dispersion of $\sigma(M_v) = 0.36$ mag.
Moreover, the linearity of the Hubble diagram indicated that these SNIa
were not significantly affected by peculiar motions (after correction for
Virgo infall) or luminosity selection effects. The mean absolute magnitude
of these SNIa was found to be ${\overline{M_v({\rm{max}}}}) 
= -19.47 + 5 \log (H_0/50)$.
In order to estimate $H_0$ one must therefore determine independently the
distance to one or more SNIa host galaxy.

HST has measured distances to IC4182 (host of SN1937C) and NGC5253
(host of SN1895B and SN1972E) from observations of Cepheid variables. These
data yielded $H_0 = 52 \pm 9$, from SN1937C alone (Saha et al. 1994) and
$H_0 = 54 \pm 8$, from the average of the three SNIa (Saha et al. 1995).
In each case the SNIa were assumed to lie at the mean of the luminosity
function. More recently Riess, Press \& Kirshner (1995) have used 
the shape of the light
curve to better constrain the luminosity at maximum light of SN1972E (the
only one of the three SNIa with sufficient quality photometry to apply their
method) and find evidence that SN1972E was significantly overluminous --
yielding $H_0 = 67 \pm 8$.

In each of these analyses the Cepheid distances were determined assuming a 
distance 
modulus of $\mu = 18.5$ for the LMC and fixing the slope of the PL
relation in V and I to that obtained from a fiducial sample of LMC Cepheids
(Madore \& Freedman 1991). This was partly an attempt to avoid {\em Malmquist
bias\/} -- i.e. a systematic error in the distance determinations due to
V-band luminosity selection effects in each HST-observed galaxy (HOG), for
which there appeared to be some evidence (Saha et al. 1994). Nevertheless,
adopting the LMC slope left the results susceptible to two further
(possibly systematic) uncertainties: sampling error due to the finite size
and different period range of the LMC and HOG Cepheids; and a possible
intrinsic difference in PL slope due to e.g. metallicity effects (c.f. Chiosi,
Wood \& Capitanio 1993).
Although allowance has been made for these sources of uncertainty in the
error budget for the published $H_0$ estimates, our aim in this work is to
{\em explicitly\/} address their impact on $H_0$ by fitting PL relations to a
composite sample of Cepheids in {\em both\/} the LMC and each HOG.

\section{Method}
Our method of calibration is essentially the same as that developed to
calibrate the Tully-Fisher relation for spiral galaxies in clusters
(c.f. Hendry et al. 1996, and references therein), where the issues of
sampling error and luminosity selection are generally of greater concern.
Full details of the method applied to Cepheids will be presented in Hendry
\& Kanbur (1996, in preparation) and we merely summarise the 
principal points here. We
assume an intrinsic absolute magnitude--log period relation which
is linear and with absolute magnitude residuals which are Gaussian with mean
zero and dispersion independent of period. We then determine the conditional
distribution of absolute magnitude, given log period, in V for
{\em observable\/} Cepheids after imposing a sharp V-band apparent
magnitude limit. We next use this conditional distribution to form a 
conditional likelihood function for the apparent magnitudes and periods
observed in the LMC and the HOG, introducing their relative distance
modulus, $\Delta \mu$, as an additional unknown parameter. We then obtain
maximum likelihood estimates of the slope and zero point of the composite
PL relation, and of $\Delta \mu$.

We applied this calibration method using mean magnitudes and periods for
the Cepheids in IC4182 and NGC5253 as published in Saha et al. (1994, 1995).
For the LMC we used `raw' magnitudes and periods from Madore (1985, Table I)
which was the main source of the calibrating sample used in Madore \&
Freedman (1991). We corrected all magnitudes for galactic (foreground)
extinction using B-band values from the Lyon Extragalactic Database
(converting to V band following Pierce \& Tully 1992). We corrected
the LMC Cepheids individually for internal extinction based on reddening
values tabulated in Martin, Warren \& Feast (1979).
Since Saha et al. (1994, 1995) find no evidence for significant internal
extinction in either NGC5253 or IC4182 we applied no correction for internal
extinction in either galaxy. We assumed a true distance
modulus of $18.5 \pm 0.1$ for the LMC, following Madore \& Freedman (1991).

\section{Results and discussion}
Tables (1) and (2) list examples of the maximum likelihood estimates obtained
for the apparent V-band distance modulus, $\mu$, of IC4182 and NGC5253
respectively, for a set of different selected period ranges,
indicated by the lower and upper limits on $\log P$ (in days) as given in
column (2). In column (3), $N_{\rm{tot}}$
denotes the total number of Cepheids in each composite sample. In all cases
a V band selection limit at $V = 25$ was applied. The error on the estimated
distance modulus was determined from Monte Carlo simulations.

The first example calibration uses essentially the same LMC stars as are 
plotted in
Figure (4) of Madore \& Freedman (1991), plus all the HOG Cepheids which are
brighter than the magnitude limit; in the second example we instead
restrict the fit
to only the period range common to both samples. The third calibration is
an example of the opposite extreme -- where the sampled period ranges in the
LMC and HOG have {\em no\/} overlap. The fourth calibration uses the full
observable period range in the HOGs and extends the LMC period range down 
to $\sim 1.5$ days, 
which is the range used in the fitted relations published in Madore \& 
Freedman (1991) 
although the shorter period Cepheids are not plotted in their Figure (4).

We see from the Tables that the estimated distance moduli are robust to the 
selected period range, even when the ranges in the LMC and HOG are disjoint 
(although of course the uncertainty on $\mu(V)$ is considerably larger in 
this case -- particularly for NGC5253) and the quoted error bands for 
$\mu(V)$ overlap. It
seems clear, therefore, that sampling error and any intrinsic differential in 
slope of the PL relation do not significantly change the estimated distance 
modulus to either galaxy. Moreover, since our results are in excellent
agreement with those of Saha et al. (1994, 1995), this confirms that the 
systematic effects of sampling error and luminosity selection in the SNIa 
host galaxies are small -- thus vindicating the distance determinations of 
Saha et al. (1994, 1995).

\begin{table}
\caption{IC4182 apparent distance moduli: composite fits}
\begin{center}\scriptsize
\begin{tabular}{cccc}
 & log period ranges & $N_{\rm{tot}}$ & $\mu(V)$\\
\tableline
1. & LMC: 0.9 - 1.8 &    &\\ 
   & HOG: 0.4 - 1.4 & 45 & $28.34 \pm 0.10$\\ 
2. & LMC: 0.9 - 1.4 &    &\\ 
   & HOG: 0.9 - 1.4 & 22 & $28.25 \pm 0.12$\\ 
3. & LMC: 1.4 - 1.8 &    &\\ 
   & HOG: 0.4 - 1.4 & 32 & $28.45 \pm 0.18$\\ 
4. & LMC: 0.2 - 1.8 &    &\\ 
   & HOG: 0.4 - 1.4 & 60 & $28.43 \pm 0.08$\\ 
\end{tabular}
\end{center}
\end{table}

\begin{table}
\caption{NGC5253 apparent distance moduli: composite fits}
\begin{center}\scriptsize
\begin{tabular}{cccc}
 & log period ranges & $N_{\rm{tot}}$ & $\mu(V)$\\
\tableline
1. & LMC: 0.9 - 1.8 &    &\\ 
   & HOG: 0.4 - 1.4 & 36 & $28.02 \pm 0.12$\\ 
2. & LMC: 0.9 - 1.4 &    &\\ 
   & HOG: 0.9 - 1.4 & 19 & $27.99 \pm 0.10$\\ 
3. & LMC: 1.4 - 1.8 &    &\\
   & HOG: 0.4 - 1.4 & 23 & $27.97 \pm 0.28$\\ 
4. & LMC: 0.2 - 1.8 &    &\\ 
   & HOG: 0.4 - 1.4 & 51 & $28.16 \pm 0.10$\\ 
\end{tabular}
\end{center}
\end{table}

Figure (1) shows the derived intrinsic linear absolute magnitude--log period 
relation in V, corrected for luminosity selection, inferred from the 
composite samples of LMC + IC4182 (solid line) and LMC + NGC5253 (dotted 
line), using
the fourth
(and largest) calibration of Tables (1) and (2). LMC Cepheids are plotted
as crosses, IC4182 Cepheids as open triangles and NGC5253 Cepheids as
open squares. We can see from this figure that the two composite
calbrations are almost completely indistiguishable, further indicating that 
we have successfully corrected for any Malmquist bias and confirming that 
residual sampling errors are small.

\begin{figure}
%\vspace{8cm}
%\centerline{\psfig{file=mahfig1.ps,width=12cm,height=8cm,angle=-90}}
\caption{Intrinsic PL relations for composite samples (see text)}
\end{figure}

\section{Conclusions: the value of $H_0$}
Using the Cepheid apparent distance moduli deduced from the fourth
example calibration -- i.e. $\mu(V) = 28.43 \pm 0.08$ to IC4182 and
$\mu(V) = 28.16 \pm 0.10$ to NGC5253 -- and assuming both SNIa to lie at the
peak of the luminosity function we estimate $H_0 = 50 \pm 9$ from SN1937C 
and $H_0 = 56 \pm 11$ from SN1972E. The error
estimate on $H_0$ is calculated by adding in quadrature the uncertainties on 
the Cepheid distance modulus and the SNIa apparent magnitude, and a further 
(conservative!) uncertainty of 0.15 mag. to allow for differential
extinction between the Cepheids and SNIa in each HOG (see Saha et al. 1994).
Note that in Saha et al. (1995) a value of $H_0 = 58 \pm 9$ was obtained from 
SN1972E alone. Thus we find that our estimates of $H_0$ are very slightly
reduced -- which ostensibly appears to be consistent with the general
trend that luminosity selection effects tend to positively bias estimates
of $H_0$. It is interesting to note, however,
that if we adopt instead the distance moduli estimated by the second
(smaller)
calibration, using the same range of periods in the LMC and HOG, then our 
estimates for $H_0$ are 54 (IC4182) and 60 (NGC5253), which are both
larger than the Saha et al. values. This comparison demonstrates that 
sampling error can have just as large an effect as selection bias on the 
value of $H_0$,
causing it to be erroneously decreased {\em or\/} increased, although the
important point of our results is that {\em both\/} effects are shown to
be very small here.

If we apply the LCS
correction of Riess et al. (1995) to SN1972E we instead find
$H_0 = 65 \pm 8$, which is also still in excellent agreement with the value 
deduced from
the Cepheid distance to M96 (Tanvir et al. 1995) and the SN type II method
of Schmidt et al. (1994) which completely by-passes the Cepheid distance scale.

Our analysis confirms that the Cepheid distances derived to these
two SNIa host galaxies appear secure (at least provided that the LMC
distance is secure) and accounting for possible sampling error and V-band
luminosity selection does not significantly change the derived distance
moduli. The outstanding uncertainty in estimating $H_0$ with
SNIa is therefore the dispersion in the SNIa luminosity function at maximum
light; this fact underlines the difficulty in making reliable statistical
conclusions from only 2 or 3 data points. The LCS method offers one solution
to this problem by reducing the dispersion, although the validity of
luminosity--LCS correlations has recently been questioned in Tammann
\& Sandage (1995). Clearly measuring more distances to SNIa host 
galaxies would be a 
better solution. Indeed, since this conference took place distances to
three more SNIa have appeared in preprints, and the issue of whether SNIa
support a long or short distance scale should soon be resolved. Whatever
the outcome, we conclude that the reliability of Cepheid distances in
determining the SNIa zero point is not in doubt.

\acknowledgments
We thank Lukas Labhardt for kindly providing the published HST Cepheid data
in electronic form.

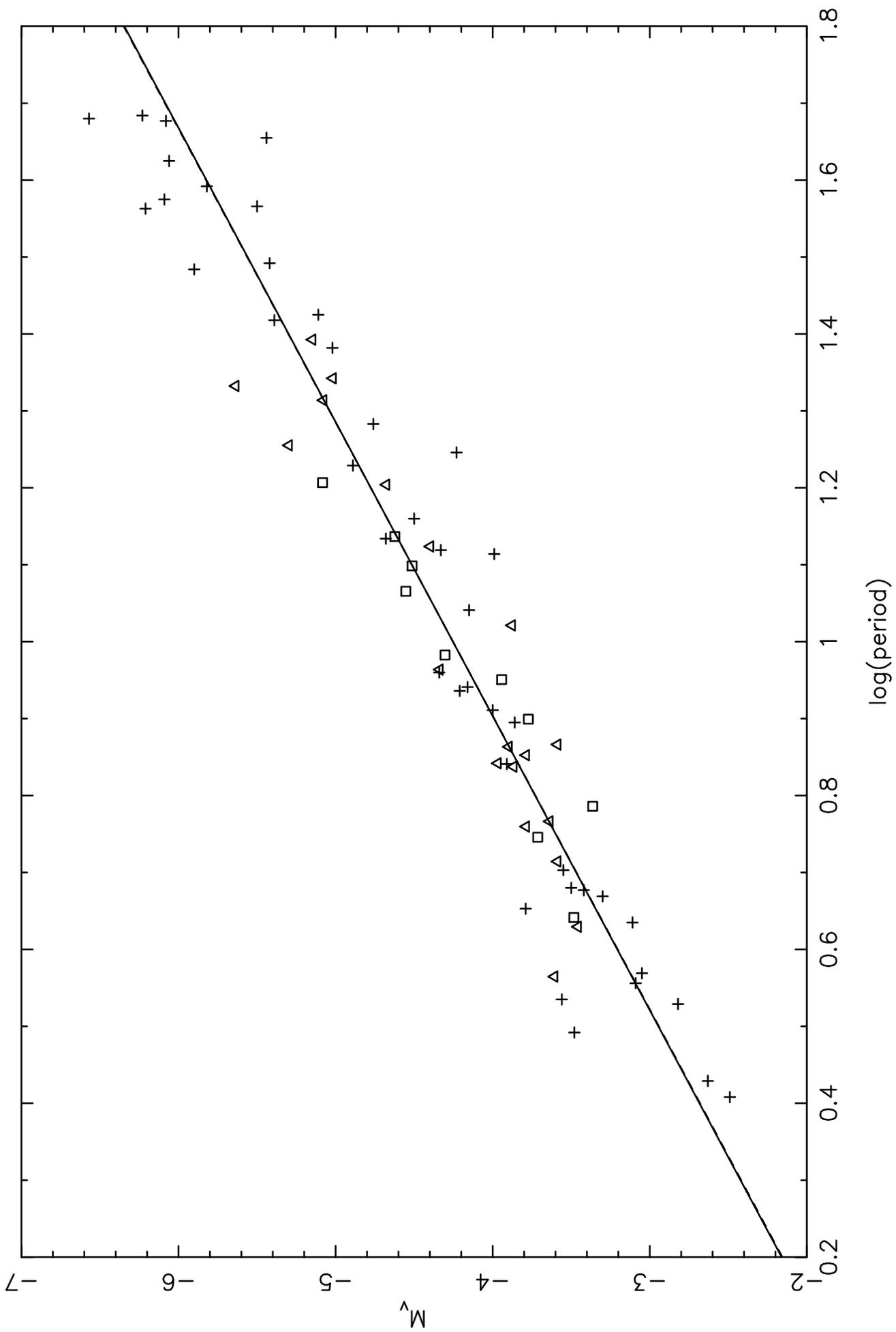
\end{document}